\documentclass[oldversion]{aa}
\usepackage{epsfig,times}
\usepackage{amssymb}
\usepackage{txfonts}
\usepackage{graphicx}
\usepackage{natbib}

\begin{document}

\title{A self-consistent model of isolated neutron stars: \\
the case of the X-ray pulsar RX J0720.4-3125}


\author{J.F. P\'erez--Azor\'{\i}n\inst{1} \and
J.A.~Pons\inst{1} \and J.A.~Miralles\inst{1} \and G. Miniutti\inst{2}}
\institute{Departament de F\'{\i}sica Aplicada, Universitat d'Alacant, 
Ap. Correus 99, 03080 Alacant, Spain \and
Institute of Astronomy , University of Cambridge,
Madingley Road, Cambridge, CB3 0HA, UK}
\date{Received...../ Accepted.....}

\abstract
{We present a unified explanation for the observed properties 
of the isolated neutron star RX J0720.4-3125 by obtaining
a self-consistent model that accounts simultaneously for
the observed X--ray spectrum and optical excess, the pulsed
fraction, the observed spectral feature around 0.3 keV, and the long--term 
spectral evolution.
We show that all observed properties are consistent
with a normal neutron star with a proper radius of about 12
km, a temperature at the magnetic pole of about 100~eV and a
magnetic field strength of $2 \times 10^{13}$~G,
value inferred from the observed period decay. The high magnetic
field produces a strong anisotropy in the surface temperature distribution.
The observed variability of the effective
temperature, strength of the spectral feature, and pulsed fraction 
are in good agreement with the predictions of our model in which the star is
subject to free precession, producing changes in the angle between the
magnetic field and the rotation axis of tens of degrees with a periodicity
of 7 years, as pointed out by other authors on the basis of phenomenological models.
 In addition to the evidence of internal toroidal components, 
we also find strong evidence of non-dipolar magnetic 
fields, since all spectral properties are better reproduced
with models with strong quadrupolar components.


}
\titlerunning{A self-consistent model of INSs: RX J0720.4-3125.}
\authorrunning{J.F. P\'erez--Azor\'{\i}n et al.}

\maketitle

\section{Introduction}

RX~J0720.4--3125 belongs to the family of radio-quiet isolated neutron
stars (INS), a puzzling group of compact objects that during the last
decade have provoked speculations about their real nature (neutron
stars or strange stars) and forced us to reconsider the thermal
emission mechanisms. RX~J0720.4--3125 was discovered with ROSAT
\citep{HMB97}, and its X--ray spectrum was soon found to be well
described by a Planckian spectrum with temperature $kT \sim 82$ eV ($k$ being
the Boltzmann constant). Similarly to the rest of INS, it is a nearby
object ($\approx 300$ pc, Kaplan et al. 2003) and shows low
interstellar absorption ($n_H = 1$--$1.5 \times 10^{20}$
cm$^{-2}$).  More interestingly, it is a confirmed X-ray pulsar with a
period of 8.391 s \citep{HMB97} and it is one of the two INSs with a
reliable measure of the period derivative $\dot{P} = 7 \times 10^{-14}
s s^{-1}$ \citep{KK05a}, which implies a magnetic field of about $B =
2.4\times10^{13}$~G \footnote{The other object with a measure of the period
derivative is RBS~1223, with $\dot{P} = 1.12 \times
10^{-13} s s^{-1}$, which implies $B = 3.4\times10^{13}$~G
\citep{KK05b}}.
Another important observational property, common to other INS, is that
the observed optical flux is larger (about a factor of 6) than
the extrapolation to the optical band of the best blackbody (BB) fit
to the X-ray emission.
This apparent optical excess flux of INSs, first observed in
RX~J1865--3754 can be explained with the existence of large temperature
anisotropies over the surface \citep{Pons02}.  In the case of
RX~J0720.4--3125, the evidence of anisotropic temperature
distributions is strongly supported by the observed X-ray pulsations
with relatively large amplitude ($\sim 11\%$).  

More recently, the story of INSs has suffered a new twist when
observations with XMM-Newton have revealed deviations from a pure BB
spectrum in the form of absorption features observed in the $0.1-1.0$
keV band.  In the particular case of RX~J0720.4--3125, a phase
dependent absorption line around 270~eV has been recently reported
\citep{Hab04}.  This feature is normally associated with proton
cyclotron resonant absorption and/or bound-bound transitions in H or
H-like He \citep{Hab03,vK04}. Both require a magnetic field of $\sim
5 \times 10^{13}$~G, consistent within with the dipole breaking
estimate of $2.4\times10^{13}$~G, but only within a factor 2.
Moreover, this estimate only reflects the large scale magnetic field
structure, whereas the presence of smaller scale inner magnetic fields
(i.e.  strong toroidal components) seems very likely because MHD core collapse
simulations show that toroidal magnetic fields are quickly generated by differential
rotation and early stage convective motions can also create toroidal fields.
Strong magnetic fields can affect
the emission properties of the neutron star surface in multiple ways.
For example, they can induce a phase transition turning the gaseous
atmosphere into a liquid or solid state \citep{Lai01}, which results
in a spectrum that, besides a reduction in the emissivity compared to a BB,
shows some particular features that could explain the observations.
This scenario has been extensively studied by different authors
\citep{Bri80,TZD04,paper1,Lai05}. In addition, if the magnetic field is
high enough to induce the condensation of the atmosphere, it will also
lead to very large anisotropies on the surface temperature
distribution \citep{GKP04,paper2,GKP06}, providing an attractive
scenario to naturally explain the observed large optical excess and
pulsed fraction of some isolated neutron stars such as RX~J0720.4--3125 and
RBS~1223. The case of RX~J0720.4--3125 is also particularly interesting
because of its clear spectral evolution \citep{deV04,Hab04,Vink04}
which has been recently associated with a $\sim$7~yr precession period
of the neutron star \citep{Hab06}.

The paper is organized as follows.  In Section 2 we review the main
properties of our theoretical model (magnetic field configuration,
temperature anisotropy, etc.). In Section 3 we summarize the X--ray
observations used in this work and in Section 4 we revisit the
phenomenological models previously adopted to describe the data. 
In section 5, we discuss and motivate the
choices made to limit the parameter--space of our models, and we
present our main results, obtained by applying our realistic and
physically motivated models to the available X--ray data.
In Section 6 and 7 we discuss the evidence of precession
and the interpretation of the excess optical flux, respectively.
Finally, in Section 8, we summarize our results and discuss open
questions and caveats.

\section{The theoretical model.} 

In previous papers \citep{paper1,paper2} we have presented the results
of detailed calculations of the temperature distribution in the crust
and condensed envelopes of neutron stars in the presence of strong
magnetic fields, by obtaining axisymmetric, stationary solutions of
the heat diffusion equation with anisotropic thermal conductivities.
Having explored a variety of magnetic field strengths and
configurations, we concluded that variations in the surface
temperature of factors 2-10 are easily obtained with $B\approx
10^{13}$-$10^{14}$ G whereas the average luminosity (and therefore the
inferred effective temperature) depends only weakly on the strength of
the magnetic field.  

Nevertheless, the luminosity is drastically
affected by the field geometry, in particular by the existence of a
toroidal component. Moreover, if the magnetic field is strong enough
to induce the condensation of the surface, the condensed surface
models also predict the existence of a spectral edge that for
$B\approx 10^{13}$-$10^{14}$ falls in the range 0.2--0.5 keV and that
can be consistent with an absorption line such as that reported in RX
J~0720.4--3125 \citep{Hab04}.  We refer to the interested reader to
the previous work \citep{paper2} for details about the calculations,
and we sketch here only the main features of the magnetic field
geometry and surface temperature distribution.

The structure of relativistic stars with both
poloidal and toroidal magnetic field components has recently been studied
\citep{IS04}. It was shown that all quantities can be determined from a
stream function that satisfies the relativistic Grad-Shafranov equation.
In the linear regime (weak magnetic field, in the sense that
deformations are small), the Grad-Shafranov equation becomes simpler
and, under the assumptions of axisymmetry and
stationarity, the general interior solution of the magnetic field
has the form \citep{IS04}
\begin{equation}
\vec{B} = B_0 \left( 2 \frac{\cos \theta}{r^2} \Psi(r), -\frac{\sin \theta}{r}
          \frac{\partial\Psi(r)}{\partial r},
          \mu \frac{\sin \theta}{r} \Psi(r) \right)
\label{bform}
\end{equation}
where $\mu$ is a constant with the interpretation of the wavenumber of the
magnetic field.  In general, the $l$-component 
of the stream function must satisfy the following differential equation
\begin{equation}
\frac{d^{2} \Psi(r)}{d r^{2}} 
 + \left(\mu^2 -  \frac{l(l+1)}{r^{2}} \right) \Psi(r) = 4 \pi r^2 \rho a_0\ ~,
\label{psieq}
\end{equation}
where $\rho$ is the energy density, $a_0=a_0(r)$ is a function that depends on the boundary
conditions and we have omitted the relativistic factors for clarity. In the
case $\mu=0$ the purely poloidal case is recovered, as discussed for
example in Konno et al. \citep{KOK99} for the weak magnetic field limit, and more
thoroughly in Bonazzola \& Gourgoulhon \citep{BG96} for the fully non--linear case. For the
general case with both poloidal and toroidal components, a similar
non--linear (although Newtonian) analysis is presented in Tomimura \& Eriguchi \cite{TE05}
where Eq.~(\ref{bform}) is generalized by considering $\mu=\mu(x)$ as a
general function of $x=r \sin\theta A_\phi$, where $A_\phi$ is the
$\phi$ component of the vector potential.

It is not the purpose of this paper to review all previous
calculations about the structure of neutron stars with magnetic fields
in detail, but we must stress the more relevant issues that lead us to
chose a particular magnetic field structure.  The key point is that
under the assumptions of stationarity and axisymmetry the electric current 
(source of the magnetic field) can only have two components, namely a
purely toroidal current which produces a poloidal field, and a second
term proportional to the magnetic field, and therefore resulting in a
force-free field with both poloidal and toroidal components. Thus, the
magnetic field is a combination of a force-free component and a purely
poloidal one.  Since the deformation of neutron stars with the
magnetic fields intensity of interest ($10^{13}$--$10^{14}$~G) are
known to be fairly small, it is reasonable to assume that the field
structure can be approximated by the force-free solution (that is
$a_0=0$ in Eq.~(\ref{bform})). In this case (and in the Newtonian
limit) Eq.~(\ref{psieq}) is a form of the Riccati-Bessel equation,
which has analytical solutions \citep{paper2}.  

In this paper, we will
consider force-free solutions with $\mu$ chosen to confine the
magnetic field to the crust and outer regions, as some of the models
discussed in previous works \citep{GKP04,paper2}, where it was shown
that this crustal confined configurations lead to large anisotropies
in the surface temperature distribution. In particular, for our
neutron star models, we have considered two cases in which
$\mu=1.34$~km$^{-1}$ and $\mu=3.87$~km$^{-1}$ depending on the
considered magnetic field configuration, either purely dipolar or
quadrupole--dominated (see P\'erez-Azor\'{\i}n et al. (2006) for plots 
of the magnetic field lines).
Notice that $\mu R$ is roughly the ratio of the toroidal to the
poloidal field, while, as mentioned above, $\mu^{-1}$ has the
interpretation of the typical length scale of the magnetic field.

\begin{figure}
\resizebox{\hsize}{!}{\includegraphics{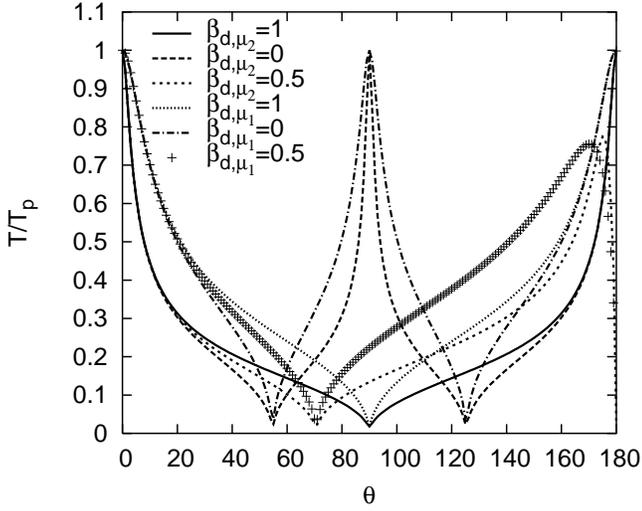}}
\caption{Surface temperature profiles as a function of the polar angle for different
magnetic field configurations according to Eq. \ref{GH}. We have taken
$\mu_1=1.34$ km$^{-1}$ and $\mu_2=3.877$ km$^{-1}$.
The different lines correspond to different relative strengths between the purely dipolar
($\beta_d=1$) and purely quadrupolar ($\beta_d=0$) models.}
\label{tdist}
\end{figure}

It turns out that for models in which the toroidal magnetic field
covers the whole crustal region (as in force-free configurations) most
of the temperature gradient is located in the envelope.  Therefore,
the classical semi-analytic temperature distribution derived by
Greenstein \& Hartke \citep{GH83}
\begin{equation}
T^{4} = T_{p}^{4} \left( \cos^{2}{\theta_{B}}
+ \frac{\kappa_{\perp}}{\kappa_{\parallel}} \sin^{2}{\theta_{B}} \right)
\label{GHf}
\end{equation}
remains a good approximation. Here $\theta_B$ is the angle 
between the magnetic field and the normal to the surface. A simple 
but accurate analytical approximation to the temperature distribution for 
dipolar force-free models, which is more realistic that discontinuous
two temperature models, is
\begin{equation}
T^4(\theta) = T_p^4 \frac{\cos^2\theta}{\cos^2\theta 
+ \frac{a+1}{4} \sin^2\theta} + T_{\rm min}^4~,
\end{equation}
where $T_p$ is the temperature at the magnetic pole, $T_{\rm min}$ is
the minimum temperature reached in the surface of the star (typically
$T_{\rm min}<10 T_p$) and the parameter $a$ takes into account the
relative strength between poloidal and toroidal components. For a
dipolar field $a=0$, while for force-free models with magnetic field
confined to the crust and envelope $a \approx 250$.  An analogous
formula when quadrupolar components are added can also be derived:
\begin{eqnarray}
T^4(\theta) &=& T_p^4 \frac{f(\theta)^2}
{f(\theta)^2 + \frac{a+1}{4}
(\beta_d + \beta_q \cos\theta)^2 \sin^2\theta}
+ T_{\rm min}^4~,
\nonumber \\
f(\theta) &=& \beta_d \cos\theta + \beta_q (3\cos^2\theta-1)/2~,
\label{GH}
\end{eqnarray}
where $\beta_d$ and $\beta_q=1-\beta_d$ are the relative weights between the
dipolar and quadrupolar components.  We have found that this approximate 
expressions are accurate to less than 3\%. 

It must be mentioned that this force-free structures can be
continuously extended into a twisted magnetosphere \citep{Lyu02, LK06},
which has been proposed to be the explanation of the infrared and
optical emission of magnetars \citep{DK05}. Within this model, the
twisted magnetic field is present not only in the magnetosphere,
but also in the envelope and probably in the crust of neutron stars.

As an example of temperature distributions, size of the hot spots, and
effect of higher order multipoles, in Fig.~\ref{tdist} we show surface
temperature profiles as a function of the polar angle for different
magnetic field configurations varying from the purely dipolar
($\beta_d=1$) to purely quadrupolar ($\beta_d=0$) models.
The main features are: i)
purely dipolar models show nearly equal antipodal hot polar caps with
only minor differences due to the Hall term in the diffusion equation
\citep{GKP06,paper2}, ii) when a quadrupolar component is present, the
north/south asymmetry is broken by the different temperatures and
sizes of the polar caps, and iii) a dominant quadrupolar component implies also the
existence of a hot equatorial belt, that can be displaced from the
equator depending on the relative strength of the dipolar and quadrupolar components. 
Of course similar but more complex geometries could be obtained by
including higher order multipoles. 
In general, smaller values of $\mu$ correspond to larger angular sizes of
polar caps and/or equatorial belts.

\section{Summary of XMM-Newton observations}

RX J~0720.4--3125 has been observed several times by {\it XMM--Newton}
and we focus here on EPIC data (pn and MOS~1) from the 8 publicly
available observations. The {\it XMM--Newton} observations span a
period of about 5.5 years starting from May 2000 (Rev.~78) to November
2005 (Rev.~1086). Here we ignore the observations performed with the
cameras operated in Small Window because of the worse calibration with
respect to Full Frame science modes. The pn data of Rev.~078 are also
excluded from the analysis due to problems in the SAS~6.5.0 reduction
pipeline (see {\tt{http://xmm.vilspa.esa.es/}} for details).  The
remaining data provide a homogeneous set and have all been collected
with the cameras operated in Full Frame mode with the Thin or Medium
filter applied and a summary of the observations is presented in
Table~\ref{XMM}. The MOS~2 data are consistent with MOS~1 and do not
add relevant information to our analysis. Photon pile--up has been
minimised by following standard procedures (see e.g. Haberl et al.
2004). Since the softest energies suffer from calibration
uncertainties, we consider the 0.18--1.2~keV band only, after having
checked that the inclusion of data down to 0.13~keV does not change
our results in any noticeable way (worsening the statistics in a
similar way for any spectral model).

In the following we present X--ray data--analysis for all observations
by using three different sets of models, a phenomenological one
(following previous works by e.g. Haberl et al. 2004; 2006) and two
realistic models obtained through detailed numerical simulations in
the theoretical framework described above. In all three cases we also
consider a sub--set of the pn--FF observations in which we force the
absorbing column density to be the same in all observations. For the
two realistic models, we also force the model normalization (related
to the NS size and distance) and the magnetic field intensity
to be the same in all observations.

\section{Phenomenological fits revised}

As already shown in previous works \citep{deV04,Hab04,Vink04}, the X--ray spectral
shape of RX J~0720.4--3125 is not that of a pure absorbed blackbody (BB).
Deviations from a thermal spectrum are seen in the 0.2--0.6~keV band
and they are well fitted by models in which a Gaussian absorption line
is added to the pure (absorbed) thermal spectrum. The line was
interpreted as cyclotron resonance scattering of protons in the NS
magnetic field \citep{Hab04}.  While the above interpretation is
certainly allowed by the data, the exact shape of the absorption
structure cannot be constrained with high confidence from the X--ray
data \citep{Vink04} and possibly describes in a phenomenological way
the deviations from a pure thermal spectrum. We point out here that BB
plus absorption line fits to the X--ray data are phenomenological
almost by definition since they cannot account simultaneously for the
observed Optical and UV data, underestimating the Optical flux by a
large factor. This is the reason why in the following we shall call
this description of the X--ray data ``phenomenological''.

We repeated this analysis by considering all
phase--averaged {\it XMM--Newton} observations and by describing the
spectra with a simple BB plus Gaussian absorption line model,
photo--electrically absorbed by a column density of gas, and our
results are reported in Table~\ref{fitf1}. Errors are given at the
90\% level for one interesting parameter. The Gaussian absorption line
width cannot be constrained with high confidence and it is fixed to
its best--fit average value to all observations (75~eV, similar to the
64~eV width imposed by Haberl et al. 2004). We point out that if the
Gaussian width is let free to vary, the absorption line energy becomes
unconstrained in many cases.  All other parameters are free to vary.
The final statistics is good for both the pn and MOS~1 cameras and the
absorption line is required at more than the 99.99 \% confidence level
in all observations except Rev.~078, 175, and 622.

As mentioned, we also consider a smaller but more homogeneous sub--set
of data from the EPIC--pn camera only, always operated in Full Frame.
In this case, since inter--calibration between different instruments
is not anymore a concern, we force the absorbing column density to be
the same in all observations, as it seems reasonable. Our results are
presented in Table~\ref{0.18-1.2BBGa} and suggest that both the BB
temperature and the absorption line equivalent width increase with
time. Both quantities increase up to Rev.~815 and decrease in the last
observation (Rev.~1086), confirming previous results \citep{Hab06}.
The temperature variation (from $\sim 85$~eV to $\sim 91$~eV) may
indicate that the hot spot(s) responsible for the X--ray pulsations of
the source has a different viewing angle (possibly due to precession), 
yielding to higher
and higher effective temperatures \cite[as proposed e.g. in][]{Vink04}.
Evidence for precession in RX~J0720.4--3125 has been indeed strengthen
in a recent work \citep{Hab06} where a $\sim$7~yr periodic variation
of the BB temperature has been claimed, based on similar fits (BB and
Gaussian absorption line) to the X--ray data. Our results on the variation
of the BB temperature are consistent with the inferred period.

\section{Spectral analysis with realistic, self--consistent models.} 

In the previous Section, we discussed simple BB plus absorption line
fits to the X--ray spectra of RX~J0720.4--3125. We remind here that
though such models do provide a very good description of the X--ray
data, they cannot explain in a self--consistent way the Optical flux
which is observed to exceed by about a factor 6 the predicted one. In
the following we present an attempt to describe the X--ray data with
synthetic spectra obtained through detailed numerical simulations
which can also simultaneously provide a good description of the
Optical/UV data.

However, the task of fitting observational data in a multi--parameter
space is numerically time--consuming (because it requires computing a
large number of models to cover a sufficient number of grid points)
and theoretically complex (because it is difficult to avoid falling in
local minima in a highly dimensional parameter space).  For this
reason we have not attempted to solve the problem using brute force
but rather to discriminate which parameters are more relevant for each
observational fact. We have considered a fiducial NS model fixing the
mass at $1.4~M_\odot$ and the radius at $12.27$~km. While the first
choice is well motivated by observations of NSs in binary systems
the radius of NSs is much less certain as
revealed by our ignorance of the details of the inner nuclear
structure and equation of state. 

In the present case, however, we are interested in the X--ray (and
Optical) emission from the star surface which, in absence of spectral
features that can reveal the redshift at the surface, will depend very
little on the assumed radius. Indeed, our models are not significantly
affected if the NS radius is varied in a sensible range (e.g.
10--15~km). This motivates and justifies the choice of a fiducial
model which greatly simplifies the numerical task of computing large
grids of models if all parameters are allowed to vary. The magnetic
field intensity is defined at the pole and we consider variations in
the range of $0.5-6 \times 10^{13}$~G, as suggested by the value of
$2.4\times 10^{13}$~G estimated from the period decay. As for the
magnetic field configuration, in the following we will explore two
different cases, both confined in the outermost regions (crust and envelope):
a purely dipolar magnetic field with $\mu = 1.34$~km$^{-1}$, and a
quadrupole--dominated one ($\beta_d = 0.05$) with $\mu = 3.87$~km$^{-1}$. 
This is only a small sub--set of
the possible configurations which has been chosen after a relatively
time--consuming assessment of which configuration provides a better
description of the X--ray data. In particular, (very) different values
for the radial length scale of the magnetic field ($\mu$) resulted in
inconsistent results in which the radius of the NS, as inferred from
the X--ray spectral fits, was different from the fiducial model one.

In the future, we plan to build a much more extensive set of grid
models which will be available for X--ray spectral fitting purposes
with most of the parameters free to vary. This will result inevitably
in a better statistical description of the X--ray data than possible
at present. We consider the present work as a first attempt to
reproduce the available observational aspects with a realistic
self--consistent model, without claiming that the proposed model(s)
is (are) the best--possible solution. However, as discussed later, the results inferred from
our analysis  provide several insights and suggest the direction to
take with the goal of improving further the theoretical models and our
understanding of INSs in general.

\begin{figure}
\resizebox{\hsize}{!}{\includegraphics{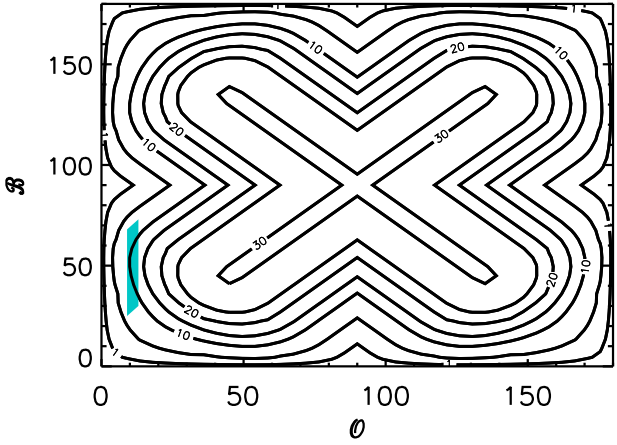}}
\caption{Contour plots for a purely dipolar magnetic field, with
  $n_{H}= 1.33 \times 10^{20}$ cm$^{-2}$, $T_{p}=115$ eV and
  $B_{p}=1.8 \times10^{13}$ G, of the pulsed fraction for a realistic
  surface temperature distribution.  ${\cal B}$ is the angle between
  rotation and magnetic axis, and ${\cal O}$ the angle between the
  rotation axis and the observer direction.  The flux has been
  obtained by integration over the whole surface taking into account
  General Relativistic light bending effects.}
\label{pfbb}
\end{figure}

Having set up a baseline model, we have built tabular XSPEC models
\footnote{Tabular models are available upon request to the authors} as
a function of the pole temperature and magnetic field strength
for several orientations within the range allowed by the pulsation
profiles, as discussed below. We then consider all the public
available {\it XMM--Newton} observations of RX J~0720.4--3125 and
compare our realistic (and limited in parameter--space) models to the
data and to the more phenomenological description of the X--ray
spectra discussed in the previous Section.

\subsection{Purely dipolar magnetic fields.}

We begin our analysis with a purely dipolar magnetic field
configuration.  This is mainly suggested by the observed regular
sinusoidal shape of the light curve of RX~J0720.4--3125 (when folded
at the spin period of 8.391~s). In the framework of our models, the
spectral shape and time--evolution mainly depends on the relative
orientation between the magnetic axis, the rotation axis, and the
observer line of sight. These can be constrained by using the observed
pulsation profile and amplitude of RX~J0720.4--3125 as detailed below.
In Fig.~\ref{pfbb} we show contour plots of the pulsed fraction, for
one of our calculations, as a function of the angle between rotation
and magnetic axis (${\cal B}$) and between the rotation and observer
direction (${\cal O}$).

First, we can classify the light curves in two groups. The models with
large ${\cal O + B}$ are characterized by a non sinusoidal pulsation,
or even two visible maxima in the pulse profile; this is because we
actually can see both poles in each period if ${\cal O + B}
>90^\circ$. Therefore, INSs which exhibit non--sinusoidal pulse
profiles do lie in this region.  On the other hand, the models with
small ${\cal O + B}$ always show one single peak in the pulse profile
which is very close to sinusoidal when either one of the angles is
small.  This is probably the case of RX~J0720.4--3125, that shows a
very regular pulsation profile.  

Next, we can reduce the range of angles to those consistent with the
observed pulsed fraction.  For a nearly spherical neutron star, the
rotation axis is essentially aligned with the (vector) total angular
momentum of the star, which is conserved. Therefore, the variation of
the angle ${\cal O}$ with time is too small to be observable,
but the star wobbles
around its symmetry axis (in general different from both, rotation and
magnetic axis) with a free precession timescale of a few years for
oblateness of the order of $10^{-7}$ \citep{Jon01}.  We have tried
different orientations in the range of ${\cal O}$ and ${\cal B}$
allowed by the observed pulsed fraction. We find that the observed pulsed
fraction values are best reproduced if ${\cal O} = 12^{\circ}$ and
${\cal B}$ varies in the range 30$^\circ$--60$^\circ$ (i.e. $180^\circ -
{\cal B}$ in the range 120$^\circ$--150$^\circ$). With this choice of
parameters, the pulsed fraction of our models always lies in the range
between 9\% and 13\%, consistent with the observed values.  Therefore,
we will focus on the vertical shaded region, fixing ${\cal O}$ and
allowing for variations in ${\cal B}$. We have build tabular XSPEC
spectral models by fixing ${\cal O} = 12^{\circ}$ with the following
free parameters: pole temperature $KT_{\rm pole}$, magnetic field
intensity $B_{\rm p}$, magnetic field orientation ${\cal B}$, and
model normalization $R_\infty /D_{300}$.  The model is applied to all
available data--sets with the addition of photo--electric absorption.

As for the phenomenological model described above, we first use all
the pn and MOS~1 observations with all parameters free to vary. Our
results are presented in Table~\ref{fitdip1}. The (unredshifted) polar 
temperature varies approximately from 110~eV to 120~eV about 25\% larger 
than the BB temperature of Table
\ref{fitf1}, while the average effective temperature over the NS
surface is about 35-41~eV, more than a factor 2 smaller than the
temperature inferred from BB fits. 
This is due to the very large
anisotropy over the surface temperature (a factor 10 between pole
and equator) induced by the magnetic field.  The X-ray spectrum is
dominated by the small hot polar area, while the extended cooler
equatorial belt is responsible for most of the optical flux which is
well reproduced (see Section~5.5).  

The statistical quality of the
fits obtained with the realistic models is comparable to that obtained
with BB plus Gaussian models (see Table~\ref{fitf1} for comparison).
However, the accuracy on the most relevant parameters (${\cal B}$,
$KT_{\rm pole}$, and $B_{\rm p}$) is limited by the fact that the
inferred variations of $n_{\rm H}$ drive the fitting results, strongly
affecting the soft energy band where most of the photons are
collected. Given the high proper motion of the source a variation in
the absorbing column cannot be excluded, but the observed variation
seem to occur almost randomly and the pn and MOS~1 data often give
inconsistent results for the same observation. It is much more
realistic to assume that the column density is the same in all
observations.

To overcome the inter--calibration uncertainties, we consider the
pn--FF observations only. As for the BB plus Gaussian fits described
above, we force the absorbing column density to be the same in all
observations. We also force the model normalization (directly related
to the NS radius and distance) and the magnetic field intensity to be
the same. Our results are presented in Table~\ref{fitdip2}. As it can
be seen by comparing with the results in Table~\ref{0.18-1.2BBGa}, the
statistical quality of the fits is worse than in the BB plus Gaussian
case. The best--fitting magnetic field intensity $B_{\rm p}$ is of the
order of $1.4\times 10^{13}$~G, about a factor 2 smaller than that
inferred from the observed period decay, possibly indicating that our
model is not an extremely accurate description of the magnetic field
geometry.

Despite the above problem, a clear long--term evolution of the
magnetic field orientation ${\cal B}$ emerges from our results. $\cal
B$ increases with time up to Rev.~815, while it
decreases in the last observation (Rev.~1086). By considering the
vertical shaded area in Fig.~\ref{pfbb} (or its equivalent
complementary at $\pi - {\cal B}$), this solution suggests that the
pulsed fraction must exhibit a long term evolution, with a maximum
pulsed fraction around ${\cal B}\sim 55^\circ$, corresponding
to the latest observations. We point out that the ${\cal B}$
long--term evolution has the same behaviour shown by the BB
temperature and the absorption line EW (see Table~\ref{0.18-1.2BBGa}),
and (see below) pulsed fraction. Our realistic model is thus
trying to reproduce the observed variations with a change in the
magnetic field orientation, which is what is expected if precession is
responsible for the long--term variability.

The model explored so far
seems to be able to describe the bulk of the observed long--term
variability in terms of precession and is therefore very promising.
However, the quality of the fits is not very satisfactory and we thus
explore in the following another case in which a strong quadrupolar
component is added to the magnetic field
configuration.

\begin{figure}
\resizebox{\hsize}{!}{\includegraphics{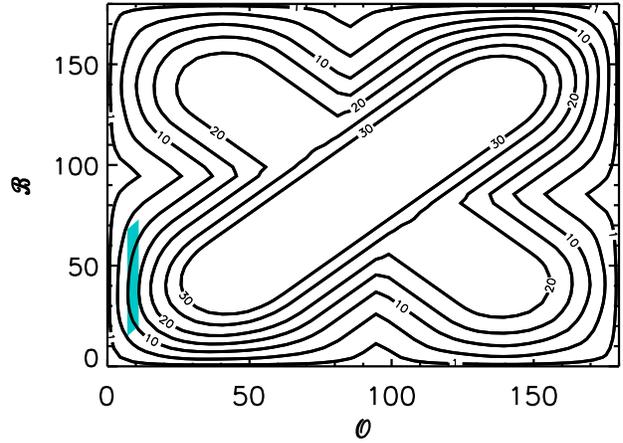}}
\caption{Contour plots of the pulsed fraction for a realistic surface
  temperature distribution produced by a quadrupole dominated magnetic
  field ($\beta_d=0.05$).  The parameters of the model are $T_p=115$
  eV ,$B_{p}= 1.8\times10^{13}$ G and $n_{H}=1.2\times10^{20}$
  cm$^{-2}$.  The flux has been obtained by integration over the whole
  surface taking into account General Relativistic light bending
  effects.}
\label{pfbbT10}
\end{figure}

\subsection{Quadrupole--dominated magnetic fields.}

In Fig.~\ref{pfbbT10} we show contour plots of the pulsed fraction for
one of the models with a dominant quadrupolar component
($\beta_d=0.05$).  The very different surface temperature
distribution (two hot spots plus a hot belt close to the equator, as
discussed in Section~2)
produces significant differences with respect to the purely dipolar
case. First, the allowed maximum pulsed fraction is much higher, up to
45\%. Second, the north/south symmetry has been broken, and the
results are different when changing either ${\cal B}$ or ${\cal O}$ by
$\pi-{\cal B}$ or $\pi-{\cal O}$ (the symmetry with respect the
simultaneous interchange of both angles by their supplementary is
kept).  Third, a portion of the hot belt will now be visible, which will
have consequences on the spectrum. We proceed as in the previous case,
and we first localize the region in which we are allowed to vary the
angles ${\cal O}$ and ${\cal B}$, keeping the pulsed fraction at about
11\%, and close to a sinusoidal shape, as observed. This
corresponds to ${\cal O}=11^\circ$ and ${\cal B} = 20^\circ-65^\circ$
(see Fig.~\ref{pfbbT10}). As mentioned, ${\cal O}=169^\circ$ and ${\cal
  B} = 115^\circ-160^\circ$ produces the same results.

We proceed as above, first applying the model to all observations and
letting all parameters free to vary, and then considering the sub--set
of pn--FF data only, forcing the absorbing column density, the model
normalization, and the magnetic field intensity to be one and the same
in all observations. Our results are presented in Table~\ref{qfit1a}
and Table~\ref{qfit1} respectively. When all parameters are free to
vary, the statistical quality of the fit is comparable to that
obtained with the phenomenological and purely dipolar--field models
(Tables~\ref{fitf1} and \ref{fitdip1}) but, as with the purely dipolar
model, the variation of the absorbing column density is a possible
sign of a degeneracy in the parameters.

We thus regard the pn--FF observations in which the parameters that
should remain constant have been forced to be the same in all
observations as the most reliable sub--set to infer information on the
remaining  model parameters (Table~\ref{qfit1}). In this case, we
obtain a very significant improvement with respect to the
purely--dipolar model and we approach the statistical quality of the
phenomenological fits. With the addition of the quadrupolar component,
the best--fitting magnetic field intensity raises to $1.8\pm 0.2
\times 10^{13}$~G, much closer to the value of $\sim 2.4 \times
10^{13}$~G inferred from the period decay. As for the dipolar model
discussed above, the magnetic field orientation ${\cal B}$ clearly
shows a long--term evolution similar to the $\pi - {\cal B}$ evolution
of the dipolar model. The issue of the long--term evolution of the
magnetic axis orientation will be discussed later in more detail in
the framework of the precession interpretation for the variability of
RX~J0720.4--3125.

\subsection{The hardness ratio anti--correlation}

The better statistical quality of the fits to the X--ray data and the
more consistent value of the magnetic field intensity are, up to some
extent, an indication of the presence of higher order multipolar
components. 
However, as recently discussed thoroughly in \cite{ZT06} there is a more 
conclusive observational fact
that allows us to distinguish between the purely dipolar and the
multipolar case. A striking feature of
the X-ray light curves of RX~J0720.4--3125 is the clear
anti--correlation of the hardness ratio with the pulse profiles in
both the hard and the soft band. This feature cannot be explained
with axisymmetric models that have also a north/south symmetry.  A
possible explanation that has been proposed is a model with two hot
spots with one of them displaced by about $20^\circ$ from the south
pole \cite{Hab06}. 

\begin{figure}
\centering
\includegraphics[width=8cm]{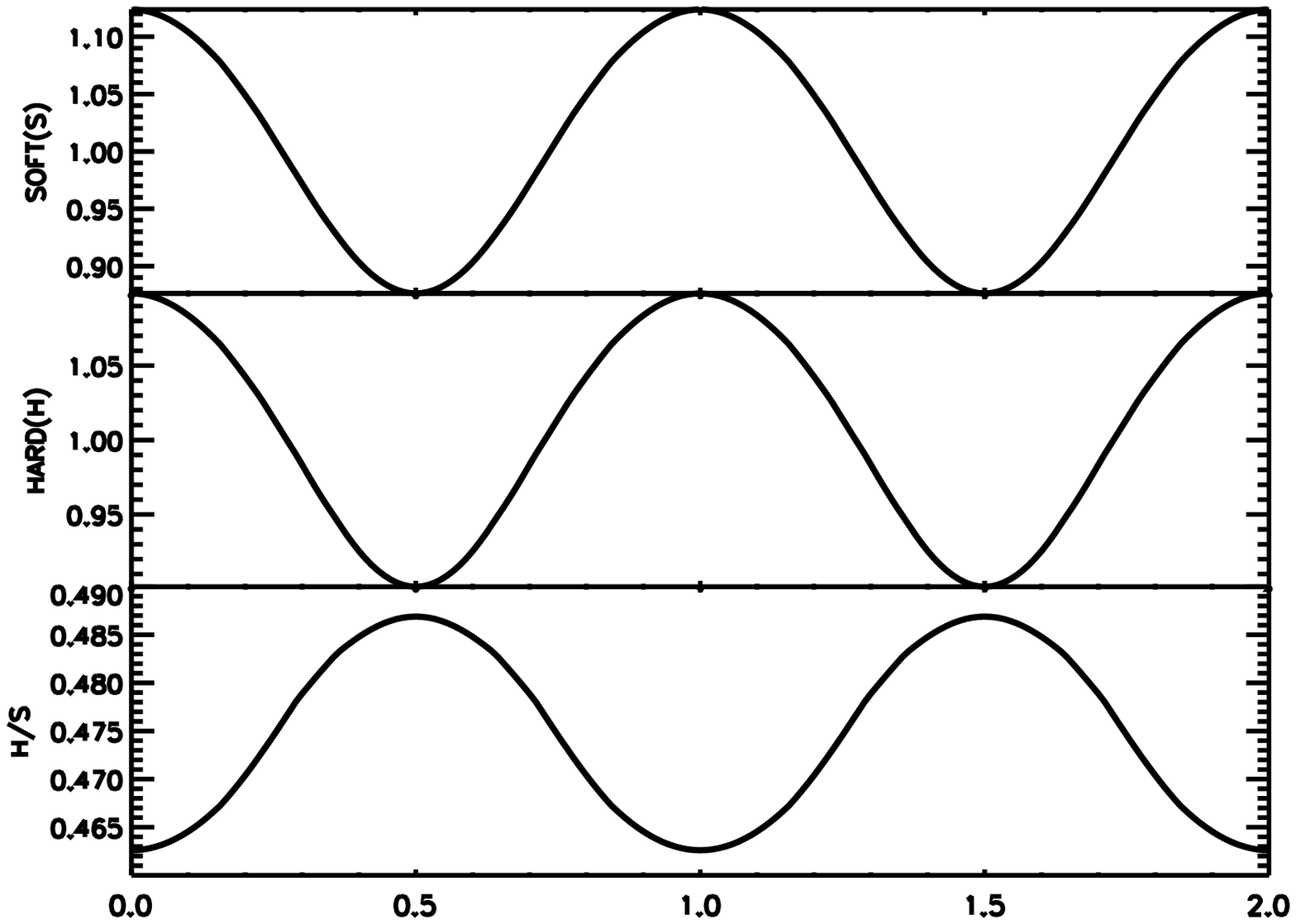}
\includegraphics[width=5cm,angle=270]{NEWFIG/fig4b.ps}
\caption{Pulse profiles in two energy bands (soft:0.12-0.4 keV,
  hard:0.4-1.0 keV) for a realistic model with ${\cal O}=11$, ${\cal
    B}=54$, $T_p=115$ eV ,$B_{p}= 1.8\times10^{13}$ G,
  $n_{H}=1.2\times10^{20}$ cm$^{-2}$ and $\beta_d=0.05$. The third
  panel is the hardness ratio (HARD/SOFT).  The bottom panel
  shows the observational data corresponding to observation rev. 534. 
  Notice that we have not attempted to perform a 
  fit of the pulse profiles, it is simply a comparison with the model
  obtained from spectral fitting. In this figure we have taken
  into account the response matrix of the instrument and the
  absorption.}
\label{LC}
\end{figure}

Motivated by the well established hardness ratio anti--correlation, we
explored whether our models can reproduce the observed behaviour.  In
Fig.~\ref{LC} we show the folded light curves in the hard and soft
bands, and the hardness ratio for one of the quadrupolar--dominated
models used in fitting the X--ray data. The observed anti--correlation
of the hardness ratio with the pulse profiles is qualitatively very
well reproduced by our model. We point out that such good agreement is
impossible to obtain with pure dipolar fields in which the North/South
symmetry is not broken. We have not attempted at this stage to find
the model that best fits the light curves and the phase averaged
spectra at the same time, which is a formidable numerical task,
especially when dealing with realistic models that (each one) need
 computational effort to be produced.  However, the good
qualitative agreement is indicative that we could be seeing a
neutron star with a {\it hot spot and a hot belt}, as predicted by
quadrupole--dominated models, as an alternative to the case of a displaced
hot spot.

If the interpretation that all variability is due to free precession
of the neutron star is correct, one should be able to find a unique
model (temperature, magnetic field configuration, etc.) that explains
all observations by only varying the relative orientation and, in our opinion,
this model approaches that solution.  Our results point in
the direction that most of the variation is actually explained by
precession of the neutron star. Notice that a precession timescale of a 
few years has also already been reported for some pulsars \citep{Link01}.

\subsection{Precession}

\begin{figure*}
\centering
\includegraphics[width=8.5cm]{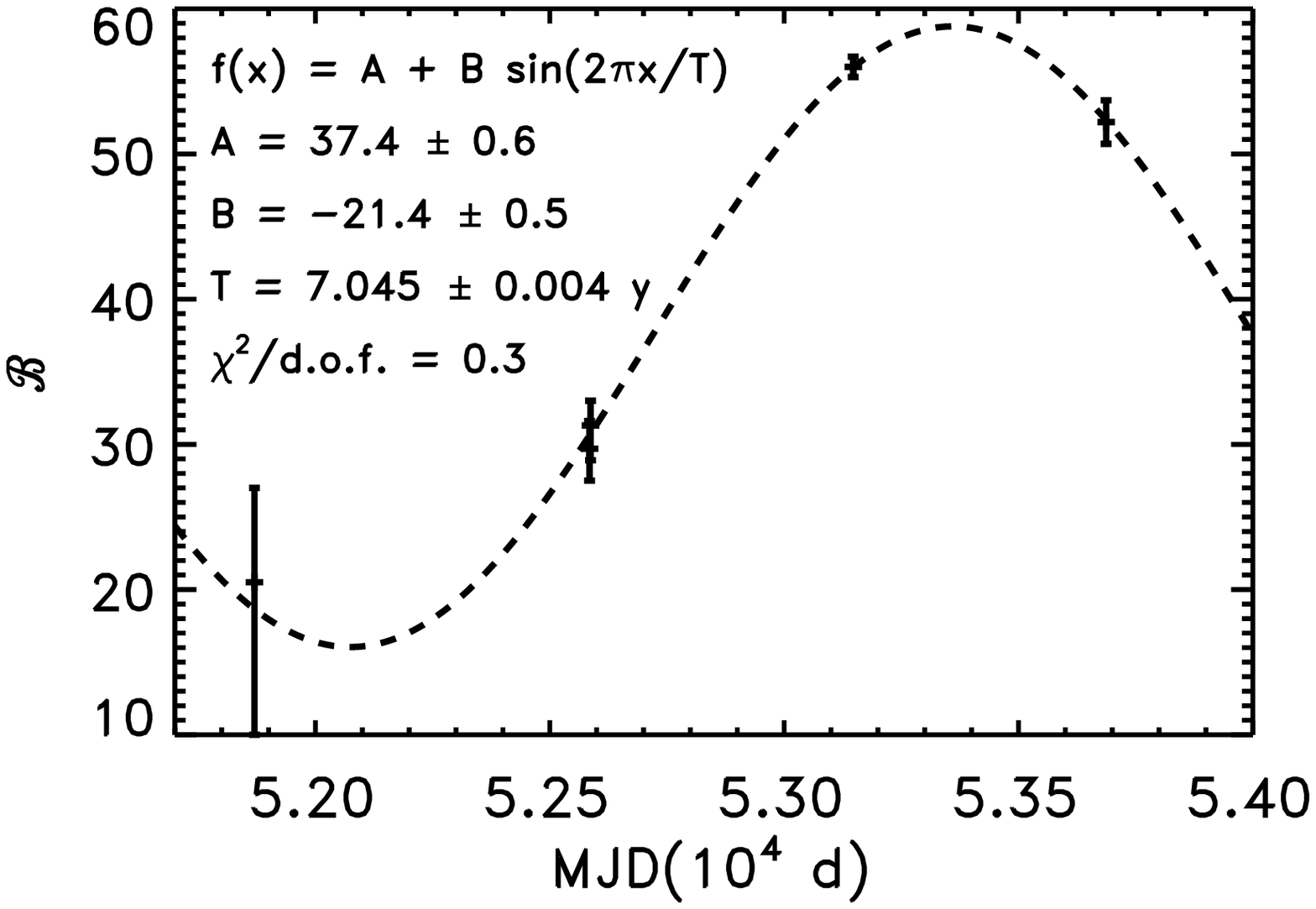}
\includegraphics[width=8.5cm]{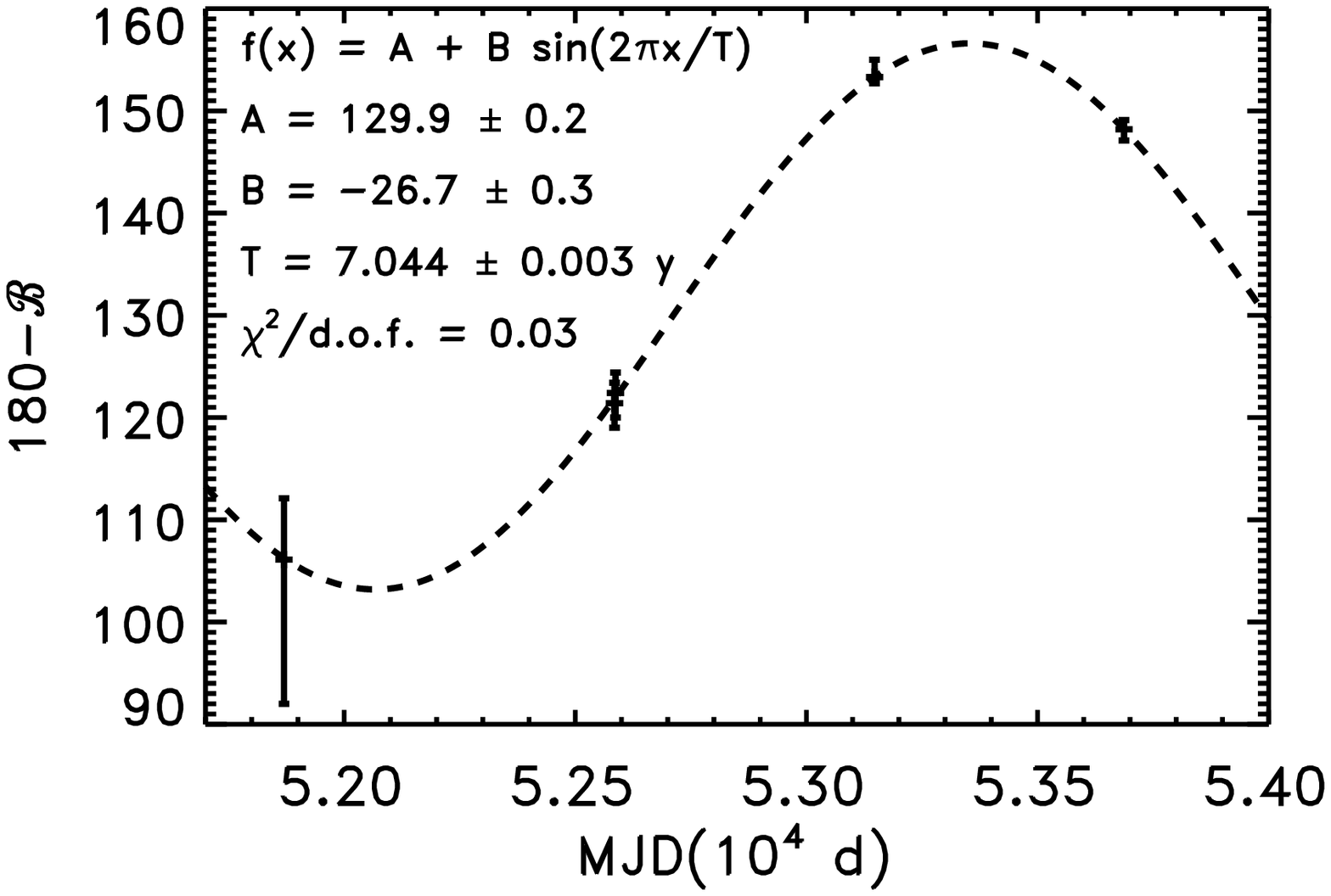}
\caption{Time variation of the orientation of the magnetic axis, ${\cal B}$, 
from the results shown in Tables \ref{fitdip2}
(left panel, $\beta_d=1$) and \ref{qfit1} (right panel, $\beta_d=0.05$). 
The solid line shows the best fit to the data (see text in figure).}
\label{prec_dp}
\end{figure*}

As mentioned, evidence for precession in RX~J0720.4--3125 has been
recently reported by Haberl et al. (2006). The evidence was based on
results from BB plus absorption line fits to the available X--ray data
in which the BB temperature shows a long--term evolution best
described by a sine wave with a period of $7.1\pm 0.5$~yr. 
We consider our results for the dipolar and quadrupole--dominated models
(see Table~\ref{fitdip2} and \ref{qfit1}) and try to fit the time
evolution of the angle ${\cal B}$ which is the relevant quantity as
far as precession is concerned. 
The results are shown in Fig.~\ref{prec_dp}.  Notice that the right panel 
shows $\pi-{\cal B}$ for better comparison. This is just a matter
of the arbitrary choice of coordinate origin of the polar angle. 
Both are consistent with the
interpretation of a precessing neutron star with a precession period
of 7 years and a relatively large wobbling angle of
$\approx 20^\circ$. Both the dipolar and quadrupole--dominated cases
are very well described by a sine wave but, as mentioned earlier, the
hardness ratio anti--correlation strongly favors model with a
significant quadrupole component breaking the north/south symmetry.
Our results are consistent with the conclusions drawn in \cite{Hab06}
with phenomenological models.

\subsection{Pulsed fraction long--term evolution}
If precession is the right interpretation fro the long--term spectral
evolution of RX~J0720.4--3125, the observed pulsed fraction should also show
long timescale variability associated with the variation of the hot
spot(s) effective area. We have then extracted 0.12--1.2~keV light
curves for the pn observations used above (the Full Frame ones),
computed the period of the observed pulsations (always consistent with
8.391~s), and obtained the pulsed fraction for each observation. Our
results suggest that the pulsed fraction is not constant in time, but
increased from 10\% to 12\% on the few years timescale spanned by the
observations. We measure a pulsed fraction of $10.1\pm0.5\%$ in
Rev.~175, $11.1\pm 0.4 \%$ in Rev.~533 and 534, and $11.7 \pm 0.5 \%$
in Rev.~815 and $11.9 \pm 0.5 \%$ in Rev.~1086. The beahaviour is
suggestive of a correlation between the BB temperature, the
absorption line equivalent width (or deviations from the BB spectrum)
and the pulsed fraction, as expected in the precession model. 

\begin{figure}
\resizebox{\hsize}{!}{\includegraphics[angle=0]{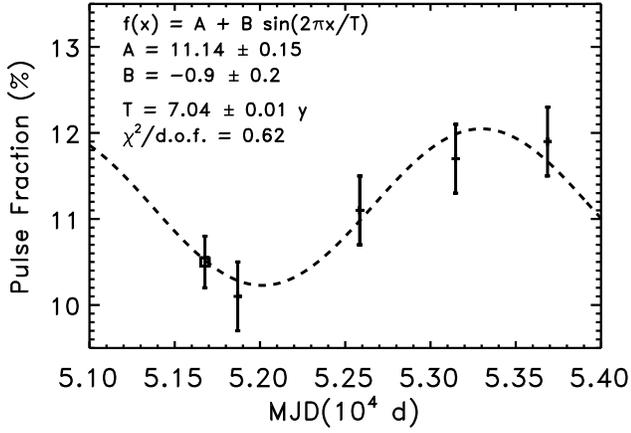}}
\caption{Variation of the pulsed fraction with time compared with a sinusoidal
profile with a periodicity of 7 years.}
\label{prec_pf}
\end{figure}

In Fig.~\ref{prec_pf} we show the
pulsed fraction long--term evolution with the addition of a first data
point from Rev~78 \citep{Hab04}. The solid line represents a fit with
a sine wave in which the period is fixed at 7~yr, while the amplitude
is a free parameter. We measure a modulation amplitude of  $1.0\pm 0.3
\%$. Although the quality of the data is not high enough to claim a
periodic evolution of the pulsed fraction, the obvious consistency
with a period of 7~yr (see Fig.~\ref{prec_pf}) strongly supports the
precession model. We point out that the pulsed fraction long--term
evolution is independent of the adopted spectral model and of the
instrument calibrations and represents thereby an independent indication that
precession is at work in RX~J0720.4--3125.

\section{Optical flux}

An issue under debate in the condensed surface models is related to
the optical flux. When the effect of motion of the ions is neglected
the optical flux is very much depressed \citep{TZD04,paper1}, but
ignoring the motion of ions at low energies is not justified. In fact,
a simple treatment of the motion of the ions as free particles leads to
very different results (Ginzburg, 1970).  This simple approach (to
consider them as free particles) could be approximately correct if the
surface is in a liquid state, but it is not proper if matter
is in solid state and ions are placed in a lattice. However,
considering them as totally fixed to the lattice can be as wrong as
letting them to move freely.  In reality, ions do actually move and
resonant plasmon or phonon excitations can be important. Moreover, 
the strong magnetic fields can modify the motion of the ions
(the ion cyclotron radius is much smaller than the separation 
between ions in the lattice).

The usual Drude model that describes the microscopic
response to electromagnetic fields in metals predicts that the dielectric function
$\epsilon$ at low energy behaves as
\begin{equation}
\epsilon = 1 - \frac{\omega_p^2}{\omega (\omega + i \Gamma)}
\end{equation}
where $\omega_p$ is the electron plasma frequency ($\omega_p^2 = 4 \pi
e^2 n_e/ m_e$) and $\Gamma$ is a relaxation coefficients.  From this
formula it is clear why at very low energy ($\omega \ll \omega_p$) the
absorption is strongly suppressed. 
In the other limit, assuming the motion of the ions is free because they are not
bound in a lattice, the effect of the magnetic field is to modify the dielectric function
(very schematically, and neglecting damping) as follows
\begin{equation}
\epsilon = 1 - \frac{\omega_p^2}{(\omega+\omega_{B,e}) (\omega - \omega_{B,i})}
\end{equation}
where $\omega_{B,e}$ and $\omega_{B,i}$ are the electron and ion cyclotron frequencies. 
Therefore, when $\omega<\omega_{B,i}$ the mode which had a large imaginary part
of the refraction index is not damped any more \citep{paper2}.

\begin{figure}
\resizebox{\hsize}{!}{\includegraphics{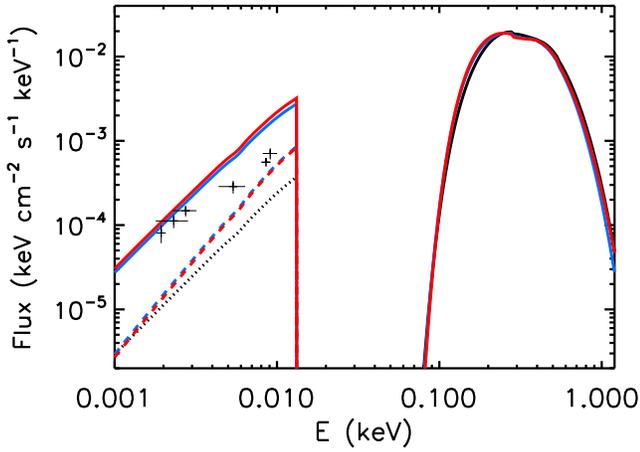}}
\caption{Spectral energy distribution of the 
best fits of the revolution 534 for phenomenological model (Table \ref{0.18-1.2BBGa}, black solid),
realistic quadrupolar dominated (Table \ref{qfit1}, blue solid), and realistic dipole 
(Table \ref{fitdip2}, red solid). 
The phase-averaged spectra of both models are very similar, and it is necessary
to analyze the spectral phase evolution to discriminate.
The dotted line is the optical tail of the BB model.
Solid  and dashed lines in the optical band correspond to models with 
free and frozen ions, respectively.
Available optical and UV data \citep{KPL03} are also shown (crosses) for comparison.}
\label{optrx}
\end{figure}

But real materials are known to
behave in a more complex way than this.  The simplest extension of the
model that can account for collective excitations of the lattice at
characteristic frequencies ($\omega_{0,j}$) is called Drude-Lorentz
model, and it consists in the combination of a Drude term with a
number of Lorentz oscillators
\begin{equation}
\epsilon = 1 - \frac{\omega_p^2}{\omega (\omega+i\Gamma)} 
+ \Sigma \frac{\omega_{p,j}^2}{(\omega_{0,j}^2-\omega^2) -i \omega \Gamma_j}~.
\end{equation}

It is clearly out of the scope of this paper to discuss the
microphysical behaviour of this condensed (liquid or solid)
atmospheres, and we do not know much about the characteristic
frequencies of the lattice and how this couples to the magnetic field. 
We just want to point out that several
important energy scales: the ion cyclotron energy $\hbar
\omega_{B,i}$, $kT$, and $\hbar \omega_{p,{\rm ion}}$ are similar 
(about 0.1 keV) under our typical
conditions and, consequently, a careful analysis of the emissivity in
the optical band is needed before we can establish severe constraints
on the models. As a compromise, since the {\it free ion} treatment
predicts an emissivity in the optical band ($\alpha=1$ corresponds to
a blackbody) of about $\alpha=0.7$ and a {\it frozen ions} model
predicts approximately $\alpha=0.2$ \citep{paper1,Lai05}, we will
show both limits in the following figures, so that reality must be
places in between this two different possibilities.

In Fig.~\ref{optrx}, we show the unfolded spectra for the best fits to
the X-ray spectrum of Rev.~534 together
with the available optical-UV data \citep{KPL03}. 
We have not attempted to fit simultaneously the optical and X-ray data, 
but the good agreement is evident.  For comparison, we also show (dotted
lines) the best BB plus Gaussian fit for the Rev.~534 data to
illustrate its similarity with the realistic model in the X--ray band
and to highlight the well known problem of under--predicting the optical
flux.  

Notice that the best fits to the optical observations at lower energies (Jul 2001)
show a somewhat higher flux than the latter optical observations (Feb 2002).
The fact the the optical observations are not consistent with a pure
Rayleigh-Jeans tail has been discussed in detail in \cite{KPL03} who
found that the best fit consists of two BBs plus a power law.
However, they also comment that the spectrum can be consistent with a
Rayleigh-Jeans tail if the deviations have a temporal nature, which is
thought to be unlikely because the X-ray flux is constant.
Interestingly, the fact that all observations fall in between the two
lines predicted by the {\it free ions} and {\it frozen ions} models may be telling us that
more effort must be placed in understanding the physics of this emission processes.

\section{Summary and conclusions} 
We have shown that realistic models of neutron stars with strong
magnetic fields are consistent with the observed X-ray and optical
spectra, the observed deviation from a pure thermal spectrum in the
X--ray band, and the long term variability. Although we do not exclude
the presence of a resonant proton cyclotron absorption or bound-bound
transitions in H, we do not need to invoke it to explain the observed
spectral edge. We stress that our model calculates the surface thermal 
emission, which could further modified by the interaction of the radiation
with the plasma in the magnetosphere.

Once the magnetic field configuration is given, we are
able to obtain a self-consistently calculated thermal spectrum that
reproduces reasonably well all available observations. Our analysis
indicates that most of the long--term spectral variation 
of the source can be explained in terms of neutron star precession. If
our interpretation is correct, we predict that the pulsed fraction,
effective temperature, and absorption line equivalent width as
obtained with phenomenological models, will decrease 
in subsequent observations of the source, until reaching similar values 
to Rev. 175.  This begins
to be apparent in the recent observation Rev. 1086, made public while
we were revising the final version of this manuscript. 
Similar conclusions have been reached independently by other groups
\citep{Hab06}.
Notice, however, that our fits still show some variability of the pole 
temperature correlated to that of the orientation angle. In theory, 
if our model were {\it perfect}, it would give the appropriate surface temperature
distribution to explain all variations keeping everything constant
except the orientation angle. We have given a step forward in that direction 
since our fits are obtained keeping fixed $n_H$, normalization (i.e. distance), 
and the star model (mass, radius, temperature distribution geometry), 
but still need to allow for small variations of temperature.
Another alternative could be that for some unknown reason 
(magnetic field evolution, diffusion) the temperature and/or size
of the hot regions has changed in a timescale of 7 years, but we are not
aware of any mechanism able to do this in such a short timescale.
If this happens to be the case, no periodicity would be expected.

Since we have only explored a few magnetic field configurations
an extensive study is very likely to improve the
statistical quality of the fits presented here. Nevertheless, we are
confident that some qualitative features are quite robust:
i) The models that reproduce the observational data have 
magnetic fields confined to the crust and the outer regions, and
strong toroidal components (about an order of magnitude larger than the 
poloidal one). This naturally leads to large surface temperature variations.
ii) Models with strong quadrupolar components are in general favored.
Higher order multipoles (octupole, etc) can in principle be added, 
but a too much patched geometry is in conflict with the large pulsation amplitudes.
iii) To explain the anti-correlation of the hardness ratio we need temperature
distributions which break the north/south asymmetry. In particular,
models with a strong quadrupole predict two hot spots with
different temperatures and sizes and a hot belt that can be shifted
from the equator.

The emission properties due exclusively to the condensed surface model
under consideration are the presence of the spectral edge at about 0.3 keV
and the absence of other spectral lines expected from gaseous atmospheres
with heavy elements. We think this part of our results is not as robust
as the strong anisotropy in the temperature distribution produced by
toroidal fields and non dipolar components. There are interesting alternatives
that can explain this spectral features equally well, but the physical
motivation of the condensed surface is that the magnetic field is high
enough to produce the condensation, if the composition is primarily iron 
or similar heavy elements. Strongly magnetized hydrogen atmospheres
(or other light elements) are still an alternative, but to our knowledge
there is no fully consistent treatment for arbitrary magnetic field orientations
and structure. More work in both lines will be certainly worth to do 
in the near future.

If future observations confirm that this INS (or others) is subject
to precession with a relatively large wobbling angle of
$\approx 20^\circ$, combined with a relatively short period of
$\sim$7~yr, it would have very interesting implications.
As pointed out by Link \citep{Link03}, 
a large wobbling angle with precession timescale of the order of 
years is in conflict with the standard picture of a type II
superconducting core. This scenario favors either type I
superconductivity or a situation in which the magnetic field does not
penetrate into the core and it is mainly confined to the crust and
outer regions. We point out that this was exactly the starting
assumption of our model, a magnetic field living in the crust needed
to explain the large surface temperature anisotropies
\citep{paper2,GKP06}. 
Alternatively, Jones \citep{Jon04} suggests that long-term precession
leads to the conclusion that nuclei and superfluid neutrons do not
coexist, which also would establish severe constraints on the equation
of state. At present, the uncertainties in our knowledge of vortex dynamics
in the inner crust can lead to different conclusions, all of them interesting
and based on the observational confirmation of long-lived under-damped precession.

We also point out that using all available constraints from the X--ray
spectra, the pulsed fraction, the pulse profiles and hardness ratios,
together with their long--term evolution seems a very promising way of
constraining the theoretical models. We plan to use all the available
constraints in the near future to relax some of the assumptions made
in this work and to explore in much more detail the vast parameter space.
We are also working to study other NSs that we expect to be well described 
by these models.  Preliminary results on RBS1223 are promising, since
we could find good fits to the very irregular light curve and the
large pulsed fraction, again with quadrupolar magnetic fields.
Another interesting NS is the pulsar PSR J1119-6127, that has a very
high pulsed fraction ($70$ \%). This is impossible to reproduce with
purely dipolar components but we found some (still axisymmetric) magnetic 
field configurations
that can explain this extremely large variability. Results about
this objects and other isolated neutron stars will be reported in
future work in preparation. In parallel, the issue of the magnetic
field stability and its diffusion timescale remains open.
Our final goal is to find a general solution pointing towards a 
unified picture in which INSs are just old, cold
magnetars whose magnetic fields are a few times smaller than usual,
maybe because they have decayed during their lifetime.

\begin{acknowledgements}
This work has been supported by the Spanish Ministerio
de Ciencia y Tecnolog\'{\i}a grant AYA 2004-08067-C03-02.
JAP is supported by a {\it Ram\'on y Cajal} contract from the Spanish MEC.
\end{acknowledgements}

\bibliographystyle{aa}


\onecolumn






\begin{table}
\caption{XMM-Newton observation information. All observations are
  performed in Full Frame with the Thin Optical filter applied except
  in Rev.~175 where the Medium filter was used.}
\begin{tabular}{lccc}
\hline
\hline\noalign{\smallskip}
{Rev.} & {Instrument}   & {Epoch} & {Exposure}\\
\hline
78 & mos1 & 2000 May 13& 48~ks\\
175 & pn  & 2000 Nov. 21--22& 23~ks\\
533 & pn  & 2002 Nov. 6--7& 26~ks\\
533 & mos1  & 2002 Nov. 6--7& 30~ks\\
534 & pn  & 2002 Nov. 8--9& 27~ks\\
534 & mos1  & 2002 Nov. 8--9& 32~ks\\
622 & mos1  & 2003 May 2--3& 24~ks\\
711 & mos1  & 2003 Oct. 27--28& 14~ks\\
815 & pn  & 2004 May  22--23& 22~ks\\
815 & mos1  & 2004 May 22--23& 26~ks\\
1086 & pn & 2005 Nov. 12--13& 34~ks\\
\hline\noalign{\smallskip}
\end{tabular} 
\label{XMM}
\end{table} 


\begin{table}
\caption{The pn and MOS~1 data for all observations are
fitted in the 0.18--1.2~keV band with an absorbed BB model
including a Gaussian absorption line.}
\begin{tabular}{l|ccccc}
\hline
\hline\noalign{\smallskip}

{Rev.--Inst.}  &  {$n_{H}$} &   {kT}   & 
{$E_{\rm{line}}$} & {$-EW_{\rm{line}}$} & {$\chi^2/dof$} \\
{} & {$10^{20}$ cm$^{-2}$} & {(eV)} & {(eV)} 
& {(eV)} & {} \\
\hline\noalign{\smallskip}
078--mos1 & $0.72^{+0.35}_{-0.33}$  & $84.4^{+1.1}_{-1.8}$
& $340^{+140}_{-165}$  & $<20$ & 58/53 \\
175--pn & $0.82^{+0.28}_{-0.42}$  & $83.7^{+0.7}_{-1.2}$ &
$249^{+102}_{-105}$  & $<40$ & 141/152 \\
533--pn & $1.20^{+0.11}_{-0.27}$       &
$85.0^{+1.1}_{-1.0}$    & $297^{+22}_{-19}$  & $32^{+11}_{-10}$ &
178/160 \\
533--mos1 & $1.20^{+0.25}_{-0.57}$       &
$87.0^{+1.9}_{-1.8}$    & $301^{+53}_{-40}$   & $35^{+25}_{-24}$ & 58/50 \\
534--pn & $1.05^{+0.19}_{-0.42}$       &
$85.7^{+1.1}_{-1.1}$    & $308^{+18}_{-22}$   & $35^{+8}_{-6}$ & 159/162 \\
534--mos1 & $0.82^{+0.31}_{-0.36}$       &
$86.5^{+1.3}_{-2.5}$    & $320^{+50}_{-35}$   & $25^{+18}_{-18}$ & 69/51 \\
622--mos1 & $1.40^{+0.32}_{-0.82}$       &
$86.5^{+2.2}_{-1.9}$    & $350^{+130}_{-150}$   & $<50$ & 46/48 \\
711--mos1 & $1.80^{+1.05}_{-1.10}$       &
$90.3^{+2.2}_{-2.8}$    & $370^{+40}_{-55}$   & $45^{+23}_{-21}$ &
47/47 \\
815--pn & $1.32^{+0.35}_{-0.19}$       &
$91.4^{+1.1}_{-1.2}$    & $308^{+11}_{-9}$   & $70^{+7}_{-9}$ & 160/163 \\
815--mos1 & $1.72^{+0.63}_{-1.05}$       &
$91.0^{+2.8}_{-2.9}$    & $370^{+20}_{-20}$   & $50^{+20}_{-15}$ &
50/52 \\
1086-pn & $1.39^{+0.30}_{-0.17}$ & 
$89.6^{+1.2}_{-0.5}$& $315^{+9}_{-10}$& $56^{+6}_{-6}$ & 201/170 \\
\\
\hline
\end{tabular} 
\label{fitf1}
\end{table} 


\begin{table}
\caption{Joint fits to all EPIC pn and MOS1 for FF mode observations in the
0.18--1.2~keV band with absorbed BB plus absorption Gaussian line
model. The absorption is forced to be the same in all
observations. NC indicates that a parameter cannot be constrained
(i.e. the corresponding model  is not required by the data).
}
\begin{tabular}{l|ccccccr}
\hline
\hline
\noalign{\smallskip}
{Rev.--Inst.} & {$n_{H}$}   & {KT}    & {$E_{line}$} & {$\sigma_{line}$} & {$-EW_{line}$} & 
{$R_{\infty}/D_{300}$} & {$\chi^{2}$/d.o.f.} \\
{} & {($10^{20}  cm^{-2}$)}   & {(eV)}    & {(eV)} &{(eV)} & {(eV)} &  {(km/300 pc)}&  \\
\hline
\\
175-pn &  1.09$^{+0.10}_{-0.10}$ & 84.45$^{+0.05}_{-0.03}$ & NC & $75^{f}$ & NC  & 5.5$^{+1.2}_{-1.0}$  & \\
533-pn &  = & 85.45$^{+0.03}_{-0.05}$ & 289$^{+11}_{-10}$& = &33.8$^{+3.0}_{-2.8}$& 5.6$^{+1.5}_{-1.4}$ & \\
534-pn &  = & 85.64$^{+0.08}_{-0.04}$ & 311$^{+10}_{-11}$& = &34.4$^{+3.1}_{-3.0}$& 5.5$^{+1.1}_{-1.3}$ & \\
815-pn &  = & 91.61$^{+0.05}_{-0.04}$ & 307$^{+5}_{-6}$  & = &69.8$^{+2.7}_{-2.8}$& 5.0$^{+1.2}_{-1.1}$ & \\
1086-pn & = & 90.51$^{+0.06}_{-0.03}$ & 301$^{+6}_{-6}$  & = &59.3$^{+2.4}_{-2.3}$& 5.1$^{+1.0}_{-1.2}$ & 902/811 \\
\\
\noalign{\smallskip}\hline\noalign{\smallskip}
\end{tabular}
\label{0.18-1.2BBGa}
\end{table}


\begin{table}
  \caption{X-ray spectral analysis.  Individual fits to the EPIC-pn and EPIC-MOS1 observations 
    in the energy band 0.18-1.2 keV for a purely 
    dipolar magnetic field structure ($\beta_d=1.0$), $\mu=1.34$ and $\mathcal{O}= 12^{\circ}$. 
    We allow to vary the magnetic field intensity at the pole ($B_p$) and the unredshifted
    temperature at the pole (k$T_{pole}$). }
\begin{tabular}{l|cccccr}
\hline
\hline\noalign{\smallskip}

{Rev.--Inst.} & {$n_{H}$}   & {${\cal B}$}    & {k$T_{pole}$} &
{$R/D_{300}$} & {$B_p$} & {$\chi^{2}$/d.o.f.} \\
{} & {($10^{20}  cm^{-2}$)}   & {}    & {(eV)} &
{(km/300 pc)} & {($10^{12}$ G)} &  {}\\

\hline\noalign{\smallskip}
78-mos1 &  0.55$^{+0.14}_{-0.20}$ & 58.0$^{+12.4}_{-8.6}$ & 118.1$^{+3.7}_{-5.1}$  & 14.1$^{+10.0}_{-9.3}$ & 8.4$^{+3.4}_{-3.3}$  & 54/53 \\
175-pn &  1.06$^{+0.11}_{-0.12}$ & 52.7$^{+18.9}_{-9.2}$ & 115.1$^{+4.6}_{-2.6}$  & 15.2$^{+20.4}_{-9.4}$ & 10.7$^{+2.5}_{-3.8}$  & 134/152 \\
533-pn &  2.15$^{+0.15}_{-0.13}$ & 57.4$^{+7.1}_{-10.8}$ & 111.5$^{+3.3}_{-2.6}$  & 19.9$^{+11.0}_{-11.2}$ & 17.4$^{+5.4}_{-3.9}$  & 175/160 \\
533-mos1 &  0.79$^{+0.96}_{-0.28}$ & 26.1$^{+35.6}_{-25.1}$ & 122.4$^{+2.0}_{-3.8}$  & 11.3$^{+15.2}_{-5.8}$ & 24.9$^{+10.2}_{-7.1}$  & 57/50 \\
534-pn &  1.91$^{+0.11}_{-0.20}$ & 50.6$^{+5.6}_{-24.1}$ & 113.3$^{+5.6}_{-2.2}$  & 17.8$^{+10.2}_{-10.4}$ & 21.5$^{+3.1}_{-7.1}$  & 157/162 \\
534-mos1 &  1.00$^{+0.35}_{-0.27}$ & 63.7$^{+4.1}_{-42.2}$ & 113.9$^{+6.9}_{-6.6}$  & 17.4$^{+11.7}_{-13.4}$ & 14.3$^{+24.2}_{-3.7}$  & 68/51 \\
622-mos1 &  2.60$^{+0.49}_{-0.32}$ & 74.8$^{+6.2}_{-10.4}$ & 108.3$^{+3.3}_{-3.2}$  & 25.3$^{+7.3}_{-7.8}$ & 15.0$^{+21.8}_{-10.0}$  & 45/48 \\
711-mos1 &  2.39$^{+0.65}_{-0.31}$ & 54.2$^{+8.5}_{-8.5}$ & 123.6$^{+2.2}_{-1.2}$  & 15.1$^{+11.5}_{-7.2}$ & 25.0$^{+9.4}_{-5.6}$  & 50/47 \\
815-pn &  2.53$^{+0.09}_{-0.09}$ & 44.1$^{+2.3}_{-1.8}$ & 124.8$^{+1.0}_{-0.5}$  & 13.9$^{+5.5}_{-5.5}$ & 25.0$^{+0.9}_{-0.5}$  & 193/163 \\
815-mos1 &  2.40$^{+0.32}_{-0.25}$ & 55.1$^{+3.5}_{-5.0}$ & 123.8$^{+2.0}_{-1.1}$  & 14.9$^{+7.3}_{-10.1}$ & 25.0$^{+3.0}_{-2.1}$  & 55/52 \\
1086-pn & 2.68$^{+0.06}_{-0.07}$ & 49.2$^{+1.3}_{-1.5}$ & 121.9$^{+0.4}_{-0.3}$  & 16.1$^{+4.3}_{-6.0}$ & 25.0$^{+0.5}_{-0.5}$  & 232/170 \\

\\
\hline

\end{tabular}
\label{fitdip1}
\end{table} 

\begin{table}
\caption{Joint fits to EPIC FF mode observations for a purely  dipolar field ($\beta_d=1$) 
with $\mu=1.34$ and  ${\cal O}=12^\circ $, forcing the hydrogen column density, 
the normalization constant and the magnetic field intensity to be the same for all observations.
}
\begin{tabular}{l|cccccr}
\hline
\hline\noalign{\smallskip}

{Rev.--Inst.} & {$n_{H}$}   & {${\cal B}$}    & {k$T_{pole}$} & {$R/D_{300}$} & {$B_p$} & {$\chi^{2}$/d.o.f.} \\
{} & {($10^{20}  cm^{-2}$)}   & {}    & {(eV)} & {(km/300 pc)} & {($10^{12}$ G)} &  {} \\

\hline\noalign{\smallskip}

175-pn &  1.33$^{+0.04}_{-0.02}$ & 20.5$^{+6.5}_{-10.5}$ & 120.6$^{+0.2}_{-0.6}$  & 11.3$^{+1.1}_{-1.8}$ & 14.1$^{+0.6}_{-0.2}$  & \\
533-pn &  = & 29.7$^{+1.9}_{-2.2}$ & 123.3$^{+0.3}_{-0.4}$  & = & = &  \\
534-pn &  = & 31.3$^{+1.7}_{-2.4}$ & 123.5$^{+0.4}_{-0.4}$  & = & = &  \\
815-pn &  = & 56.0$^{+0.7}_{-0.7}$ & 132.3$^{+0.5}_{-0.4}$  & = & = &  \\
1086-pn & = & 52.2$^{+1.5}_{-1.5}$ & 131.2$^{+0.3}_{-0.7}$  & = & = &  1570/819\\
\\
\hline
\end{tabular}
\label{fitdip2}
\end{table}

\begin{table}
\caption{Individual fits to the EPIC-pn and EPIC-MOS1 observations in the energy band
0.18-1.2 keV for a quadrupole dominated magnetic field ($\beta_d=0.05$) with $\mu=3.87$ 
and $\mathcal{O}= 11^{\circ}$. 
We allow to vary the magnetic field intensity and the temperature at the pole.}
\begin{tabular}{l|cccccr}
\hline
\hline\noalign{\smallskip}

{Rev.--Inst.} & {$n_{H}$}   & {${\cal B}$}    & {k$T_{pole}$} &
{$R/D_{300}$} & {$B_p$} & {$\chi^{2}$/d.o.f.} \\
{} & {($10^{20}  cm^{-2}$)}   & {}    & {(eV)} &
{(km/300 pc)} & {($10^{12}$ G)} &  {}\\

\hline\noalign{\smallskip}
78-mos1 &  0.43$^{+0.37}_{-0.27}$ & 56.0$^{+88.0}_{-27.1}$ & 105.6$^{+2.1}_{-1.1}$  & 12.5$^{+9.4}_{-6.0}$ & 14.2$^{+21.9}_{-9.2}$  & 56/53 \\
175-pn &  0.89$^{+0.13}_{-0.14}$ & 70.7$^{+65.0}_{-18.0}$ & 104.4$^{+0.5}_{-0.5}$  & 13.3$^{+9.7}_{-4.9}$ & 39.6$^{+11.6}_{-11.7}$  & 138/152 \\
533-pn &  1.71$^{+0.08}_{-0.10}$ & 51.2$^{+3.4}_{-6.7}$ & 104.2$^{+0.5}_{-0.17}$  & 16.2$^{+6.1}_{-5.7}$ & 54.4$^{+6.6}_{-8.4}$  & 173/160 \\
533-mos1 &  0.83$^{+0.79}_{-0.31}$ & 31.4$^{+36.4}_{-13.3}$ & 106.4$^{+1.3}_{-2.0}$  & 15.2$^{+9.9}_{-8.5}$ & 24.9$^{+20.0}_{-11.9}$  & 57/50 \\
534-pn & 1.65$^{+0.07}_{-0.14}$ & 47.4$^{+2.6}_{-7.0}$ & 104.1$^{+0.6}_{-0.3}$  & 16.4$^{+6.0}_{-5.9}$ & 55.0$^{+5.0}_{-10.4}$  & 160/162 \\
534-mos1 &  0.41$^{+0.27}_{-0.40}$ & 43.5$^{+9.6}_{-26.1}$ & 104.9$^{+1.6}_{-1.2}$  & 14.4$^{+10.0}_{-7.7}$ & 43.7$^{+11.3}_{-18.3}$  & 69/51 \\
622-mos1 &  0.44$^{+0.64}_{-0.39}$ & 59.3$^{+108.0}_{-49.6}$ & 106.6$^{+1.6}_{-1.3}$  & 11.8$^{+12.5}_{-5.7}$ & 24.9$^{+22.7}_{-19.8}$  & 45/48 \\
711-mos1 &  2.56$^{+0.39}_{-0.88}$ & 32.7$^{+6.14}_{-8.2}$ & 107.7$^{+7.5}_{-1.4}$  & 17.2$^{+11.1}_{-11.9}$ & 55.0$^{+5.0}_{-25.0}$  & 50/47 \\
815-pn &  1.99$^{+0.09}_{-0.11}$ & 25.6$^{+1.7}_{-1.6}$ & 111.9$^{+1.8}_{-0.6}$  & 15.4$^{+3.0}_{-3.1}$ & 25.0$^{+1.2}_{-0.9}$  & 206/163 \\
815-mos1 &  2.00$^{+0.27}_{-0.40}$ & 36.7$^{+3.2}_{-6.2}$ & 109.9$^{+2.5}_{-0.8}$  & 15.4$^{+9.1}_{-7.5}$ & 54.0$^{+6.0}_{-8.4}$  & 54/52 \\
1086-pn & 1.90$^{+0.03}_{-0.07}$ & 25.4$^{+1.2}_{-1.0}$ & 109.5$^{+0.5}_{-1.3}$  & 16.5$^{+3.2}_{-3.0}$ & 25.0$^{+0.9}_{-0.7}$  & 255/170 \\

\hline

\end{tabular}
\label{qfit1a}
\end{table} 


\begin{table}
\caption{Joint fits to EPIC-pn and EPIC-MOS1 FF mode observations for a 
quadrupole dominated magnetic field ($\beta_d=0.05$) with $\mu=3.87$ and $\mathcal{O}= 11^{\circ}$, 
forcing the hydrogen column density, 
the normalization constant and the magnetic field intensity to be the same for all observations.
}
\begin{tabular}{l|cccccr}
\hline
\hline\noalign{\smallskip}
{Rev.--Inst.} & {$n_{H}$}   & {${\cal B}$}    & {k$T_{pole}$} & {$R/D_{300}$} & {$B_p$} & {$\chi^{2}$/d.o.f.} \\
{} & {($10^{20}  cm^{-2}$)}   & {}    & {(eV)} & {(km/300 pc)} & {($10^{12}$ G)} &  {}\\
\hline\noalign{\smallskip}

175 &  1.23$^{+0.04}_{-0.04}$ & 73.9$^{+6.0}_{-14.1}$ & 104.6$^{+0.4}_{-0.4}$  & 12.7$^{+1.0}_{-2.1}$ & 18.2$^{+1.6}_{-1.5}$  & \\
533 &  = & 58.6$^{+2.0}_{-2.4}$ & 107.5$^{+0.2}_{-0.3}$  & = & = &  \\
534 &  = & 57.6$^{+2.0}_{-2.4}$ & 107.5$^{+0.4}_{-0.3}$  & = & = &  \\
815 &  = & 26.9$^{+1.7}_{-0.6}$ & 118.2$^{+0.5}_{-0.2}$  & = & = &  \\
1086 & = & 31.8$^{+0.9}_{-1.1}$ & 116.3$^{+0.5}_{-0.3}$  & = & = &  1192/819\\

\\

\hline
\end{tabular}
\label{qfit1}
\end{table}

\end{document}